\documentclass[a4paper]{jpconf}
\usepackage{epsfig}
\usepackage{amssymb}
\usepackage{amsthm}
\usepackage{latexsym}
\newcommand{\gtwid}{\mathrel{\raise.3ex\hbox{$>$\kern-1.1em\lower1ex\hbox{
$\sim$}}}}
\newcommand{\ltwid}{\mathrel{\raise.3ex\hbox{$<$\kern-1.1em\lower1ex\hbox{
$\sim$}}}}
\begin{document}
\title{\textbf{Generalizing Starobinski\u{\i}'s Formalism to Yukawa Theory \&
to Scalar QED}}
\author{\textbf{R. P. Woodard}}
\date{}
\address{Department of Physics, University of Florida \\
Gainesville, FL 32611-8440, USA}
\ead{\textbf{woodard@phys.ufl.edu}}
\vspace{1cm}
\begin{abstract}
During inflation quantum effects from massless, minimally coupled scalars
and gravitons can be strengthened so much that perturbation theory breaks
down. To follow the subsequent evolution one must employ a nonperturbative
resummation. Starobinski\u{\i} has developed such a technique for simple
scalar theories. I discuss recent progress in applying this technique to
more complicated models.
\end{abstract}

\section{Introduction}

On cosmological scales the universe is well described by a homogeneous, 
isotropic and spatially flat geometry,
\begin{equation}
ds^2 = -dt^2 + a^2(t) d\vec{x} \cdot d\vec{x} \; . \label{FRW}
\end{equation}
Derivatives of the scale factor $a(t)$ give the Hubble parameter 
$H(t)$ and the deceleration parameter $q(t)$,
\begin{equation}
H(t) \equiv \frac{\dot{a}}{a} \qquad , \qquad q(t) \equiv 
-\frac{a \ddot{a}}{\dot{a}^2} = -1 - \frac{\dot{H}}{H^2} \; .
\end{equation}
Inflation is defined as positive expansion ($H(t) > 0$) with negative
deceleration ($q(t) < 0$).

The homogeneity of spacetime expansion evident in (\ref{FRW}) does not change 
the fact that particles have constant wave vectors $\vec{k}$, but it does 
alter what these mean physically. In particular the energy of a particle
with mass $m$ and wave number $k = \Vert \vec{k}\Vert$ becomes time dependent,
\begin{equation}
E(t,k) = \sqrt{m^2 + \Bigl(k/a(t)\Bigr)^2} \; .
\end{equation}
This results in an interesting change in the en\-er\-gy-time uncertainty
principle which restricts how long a virtual pair of such particles
with $\pm \vec{k}$ can exist. If the pair was created at time $t$, it
can last a time $\Delta t$ given by the integral,
\begin{equation}
\int_t^{t+\Delta t} \!\!\!\!\! dt' E(t',k) \sim 1 \; .
\end{equation}
Just as in flat space, particles with the smallest masses persist 
longest. For the fully massless case the wave number factors out, leaving
an integral which can be recognized as the radius of the forward light-cone
from time $t$ to time $t + {\Delta t}$ \cite{RPW1},
\begin{equation}
\int_t^{t+\Delta t} \!\!\!\!\! dt' E(t',k) \Bigl\vert_{m=0} = 
k \int_t^{t+\Delta t} \!\! \frac{dt'}{a(t')} \simeq \frac{k}{q(t') 
H(t') a(t')} \Bigl\vert^{t+{\Delta t}}_t \; .
\end{equation}
For positive deceleration the upper limit dominates and the integral grows
without bound as ${\Delta t}$ increases. In this case the persistence
time ${\Delta t}$ is finite, although longer than in flat space. However,
for negative deceleration (inflation) it is the lower limit that
dominates, and the integral approaches a finite value as ${\Delta t}$
goes to infinity. For example, the result for de Sitter ($a(t) = e^{Ht}$
with $H$ constant) is,
\begin{equation}
k \int_t^{t+\Delta t} \!\! \frac{dt'}{a(t')} \Bigl\vert_{\rm de\ Sitter} 
= \frac{k}{H a(t)} \Bigl[1 - e^{-H \Delta t}\Bigr] \; .
\end{equation}
Therefore, any massless virtual particle which happens to emerge from the 
vacuum with $k \ltwid H(t) a(t)$ can persist forever!

Most massless particles possess conformal invariance. The change of
variables $d\eta \equiv dt/a(t)$ defines a conformal time in terms of
which the invariant element (\ref{FRW}) is just a conformal factor 
times that of flat space,
\begin{equation}
ds^2 = -dt^2 + a^2(t) d\vec{x} \cdot d\vec{x} = a^2 \Bigl(-d\eta^2
+ d\vec{x} \cdot d\vec{x}\Bigr) \; .
\end{equation}
In the $(\eta,\vec{x})$ coordinates, conformally invariant theories 
are locally identical to their flat space cousins. The rate at which 
virtual particles emerge from the vacuum per unit conformal time must 
be the same constant --- call it $\Gamma$ --- as in flat space. Hence 
the rate of emergence per unit physical time is,
\begin{equation}
\frac{dN}{dt} = \frac{dN}{d\eta} \frac{d\eta}{dt} = \frac{\Gamma}{a(t)} \; .
\end{equation}
It follows that, although any sufficiently long wavelength, massless 
and conformally invariant particle which emerges from the vacuum can
persist forever during inflation, very few will emerge.

Two kinds of massless particles do not possess conformal invariance:
minimally coupled scalars and gravitons. To see that the production of
these particles is not suppressed during inflation note that each 
polarization and wave number behaves like a harmonic oscillator,
\begin{equation}
L = \frac12 m \dot{q}^2 - \frac12 m \omega^2 q^2 \; ,
\end{equation}
with time dependent mass $m(t) = a^3(t)$ and frequency $\omega(t) = 
\frac{k}{a(t)}$. The Heisenberg equation of motion can be solved in 
terms of mode functions $u(t,k)$ and canonically normalized raising 
and lowering operators $\alpha^{\dagger}$ and $\alpha$,
\begin{equation}
\ddot{q} + 3 H \dot{q} + \frac{k^2}{a^2} q = 0 \quad \Longrightarrow \quad 
q(t) = u(t,k) \alpha + u^*(t,k) \alpha^{\dagger} \quad {\rm with} \quad
[\alpha,\alpha^{\dagger}] = 1 \; ,
\end{equation}
The mode functions $u(t,k)$ are quite complicated for a general scale 
factor $a(t)$ \cite{TW1} but they take a simple form for de Sitter,
\begin{equation}
u(t,k) = \frac{H}{\sqrt{2 k^3}} \Bigl[1 - \frac{i k}{H a(t)}\Bigr]
\exp\Bigl[\frac{ik}{Ha(t)}\Bigr] \; . \label{uform}
\end{equation}

The (co-moving) energy operator for this system is,
\begin{equation}
E(t) = \frac12 m(t) \dot{q}^2(t) + \frac12 m(t) \omega^2(t) q^2(t) \; .
\end{equation}
Owing to the time dependent mass and frequency, there are no stationary 
states for this system. At any given time the minimum eigenstate of
$E(t)$ has energy $\frac12 \omega(t)$, but which state this is changes 
for each value of time. The state $\vert \Omega \rangle$ which is 
annihilated by $\alpha$ has minimum energy in the distant past. The 
expectation value of the energy operator in this state is,
\begin{equation}
\Bigl\langle \Omega \Bigl\vert E(t) \Bigr\vert \Omega \Bigr\rangle =
\frac12 a^3(t) \vert \dot{u}(t,k) \vert^2 + \frac12 a(t) k^2
\vert u(t,k) \vert^2 \Bigl\vert_{\rm de\ Sitter} = \frac{k}{2a} + 
\frac{H^2 a}{4k} \; .
\end{equation}
If one thinks of each particle having energy $k/a(t)$, it follows that the
number of particles with any polarization and wave number $k$ grows as
the square of the inflationary scale factor,
\begin{equation}
N(t,k) = \Bigl(\frac{H a(t)}{2 k}\Bigr)^2 \; ! \label{create}
\end{equation}

Quantum field theoretic effects are driven by essentially classical 
physics operating in response to the source of virtual particles implied 
by quantization. On the basis of (\ref{create}) one might expect inflation 
to dramatically enhance quantum effects from MMC scalars and gravitons, 
and explicit studies over a quarter century have confirmed this. The 
oldest results are of course the cosmological perturbations induced by 
scalar inflatons \cite{MC} and by gravitons \cite{AAS2}. More recently it 
was shown that the one loop vacuum polarization induced by a charged MMC 
scalar in de Sitter background causes super-horizon photons to behave like 
massive particles in some ways \cite{PTW1,PTW2,PW1}. Another recent result 
is that the one loop fermion self-energy induced by a MMC Yukawa scalar in de
Sitter background reflects the generation of a nonzero fermion mass 
\cite{PW2,GP}. In the next three sections it will be explicitly shown how 
these one loop results generalize to all orders.

\section{Infrared Logarithms}

The expectation values of familiar operators typically show enhanced quantum 
effects in the form of {\it infrared logarithms}. A simple example is 
provided by the stress tensor of a massless, minimally coupled scalar 
with a quartic self-interaction,
\begin{equation}
\mathcal{L} = -\frac12 \partial_{\mu} \varphi \partial_{\nu} \varphi g^{\mu\nu}
\sqrt{-g} - \frac{\lambda}{4!} \varphi^4 \sqrt{-g} + {\rm counterterms} \; .
\end{equation}
When the expectation value of the stress tensor of this theory is computed
in de Sitter background ($a(t) = e^{Ht}$)
and renormalized so as to make quantum effects vanish at $t=0$, the results 
for the quantum-induced energy density and pressure are \cite{OW1,OW2},
\begin{eqnarray}
\rho(t) & = & \frac{\lambda H^4}{(2 \pi)^4} \Bigl\{ \frac18 
\ln^2(a) \Bigr\} + O(\lambda^2) \; , \label{rho} \\
p(t) & = & \frac{\lambda H^4}{(2 \pi)^4} \Bigl\{-\frac18 \ln^2(a) - 
\frac1{12} \ln(a) \Bigr\} + O(\lambda^2) \; . \label{pres}
\end{eqnarray}
Infrared logarithms are the factors of $\ln(a) = H t$. They arise from the
fact that inflationary particle production drives the free scalar field 
strength away from zero \cite{VF,ADL,AAS3},
\begin{equation}
\Bigl\langle \Omega \Bigl\vert \varphi^2(x) \Bigr\vert \Omega \Bigr\rangle_0 
= \frac{H^2}{4 \pi^2} \ln(a) + {\rm Divergent\ Constant} \; .
\end{equation}
This increases the vacuum energy contributed by the quartic potential, and 
the result is evident in (\ref{rho}-\ref{pres}).

Infrared logarithms arise in the one particle irreducible (1PI) functions of
this theory \cite{BOW}. They occur as well in massless, minimally coupled 
scalar quantum electrodynamics (SQED) \cite{PTW1,PTW2,PW1} and in
massless Yukawa theory \cite{PW2,GP}. The 1PI functions of pure gravity on
de Sitter background show infrared logarithms \cite{TW2,TW3}. It seems
inevitable that infrared logarithms contaminate loop corrections to the power 
spectrum of cosmological perturbations \cite{SW1,Sloth} and similar
fixed-momentum correlators \cite{SW2}. And infrared logarithms have been 
discovered in the 1PI functions and quantum-corrected mode functions of 
Einsetin + Dirac \cite{MW2,MW3}.

Infrared logarithms are fascinating because they introduce a secular element
into the usual, static expansion in the loop counting parameter. No matter
how small the coupling constant $\lambda$ is in (\ref{rho}-\ref{pres}), the
continued growth of the inflationary scale factor must eventually overwhelm 
it. When this happens, perturbation theory breaks down. For example, the 
general form of the induced energy density (\ref{rho}) is,
\begin{equation}
\rho(t) = H^4 \sum_{\ell = 2}^{\infty} \lambda^{\ell-1} \Bigl\{C^{\ell}_{0}
\ln^{2\ell-2}(a) + C^{\ell}_{1} \ln^{2\ell-3}(a) + \ldots + C^{\ell}_{2\ell-4}
\ln^2(a) \Bigr\} \; .
\end{equation}
The $C^{\ell}_{0} [\lambda \ln^2(a)]^{\ell-1}$ terms are the {\it leading 
logarithms} at $\ell$ loop order; the remaining terms are {\it subdominant 
logarithms}. Assuming that the numerical coefficients $C^{\ell}_{k}$ are of 
order one, we see that the leading infrared logarithms all become order one
at $\ln(a) \sim 1/\sqrt{\lambda}$. At this time the highest subdominant
logarithm terms are still perturbatively small ($\sim \sqrt{\lambda}$), so 
it seems reasonable to attempt to follow the nonperturbative evolution by
resuming the series of leading infrared logarithms,
\begin{equation}
\rho_{\rm lead} = H^4 \sum_{\ell = 0}^{\infty} C^{\ell}_0 \Bigl(\lambda
\ln^2(a)\Bigr)^{\ell-1} \; .
\end{equation}
This is known as the {\it leading logarithm approximation}.

\section{Starobinski\u{\i}'s Formalism for Simple Scalar Models}

Starobinski\u{\i} has long maintained that his stochastic field equations
reproduce the leading logarithm approximation \cite{AAS1}. With Yokoyama 
\cite{SY} he exploited this conjecture to explicitly solve for the 
nonperturbative, late time limit of any model of the form , 
\begin{equation}
\mathcal{L} = -\frac12 \partial_{\mu} \varphi \partial_{\nu} \varphi g^{\mu\nu}
\sqrt{-g} - V(\varphi) \sqrt{-g} \; , \label{starform}
\end{equation}
assuming only that the potential $V(\varphi)$ is bounded below. When the 
potential is unbounded below the conjecture still gives the leading infrared 
logarithms at each order, however, the theory fails to approach a static limit.

An all orders derivation has recently been given of Starobinski\u{\i}'s 
formalism \cite{RPW2,TW4}. The first step is to rewrite the operator field 
equations,
\begin{equation}
\ddot{\varphi} + 3 H \dot{\varphi} - \frac{\nabla^2}{a^2} \varphi + 
V'(\varphi) = 0 \; ,
\end{equation}
in Yang-Feldman form \cite{YF}, 
\begin{equation}
\varphi(t,\vec{x}) = \varphi_0(t,\vec{x}) - \int_0^t dt' a^{\prime 3} 
\int d^3 x' G(x;x') V'\Bigl(\varphi(x')\Bigr) \; .
\end{equation}
The free field expansion and retarded Green's function are,
\begin{eqnarray}
\varphi_0(t,\vec{x}) &\equiv & \int \frac{d^3k}{(2\pi)^3} 
\Bigl\{ u(t,k) e^{i \vec{k} \cdot \vec{x}} \alpha(\vec{k}) + u^*(t,k) 
e^{-i \vec{k} \cdot \vec{x}} \alpha^{\dagger}(\vec{k}) \Bigr\} \; , \\
G(x;x') & = & i\theta(\Delta t) \int \frac{d^3k}{(2\pi)^3} e^{i \vec{k} \cdot 
\Delta \vec{x}} \Bigl\{ u(t,k) u^*(t',k) - u^*(t,k) u(t',k) \Bigr\} \; .
\end{eqnarray}
The mode function $u(t,k)$ was given in expression (\ref{uform}). 
The canonically normalized creation and annihilation operators are 
$\alpha^{\dagger}(\vec{k})$ and $\alpha(\vec{k})$. Iterating the Yang-Feldman 
equation gives the usual perturbative expansion of the interaction picture 
field, expressed in terms of a field which is free at $t=0$.

Now consider taking the expectation value of some operator constructed from
$\varphi$, and hence from the free field $\varphi_0$. To reach leading 
logarithm order requires that every free field contributes to an infrared 
logarithm. The full result from the pairing of two free fields is,
$$\Bigl\langle \Omega_0 \Bigl\vert \varphi_0(t,\vec{x}) \varphi_0(t',\vec{x}')
\Bigr\vert \Omega_0 \Bigr\rangle = \frac1{2 \pi^2} \int dk k^2 \, u(t,k) 
u^*(t',k) \frac{\sin(k \Delta x)}{k \Delta x} \; . \eqno(25)$$
At high $k$ the mode functions and the $\sin(k \Delta x)$ oscillate, which
makes the integral converge. As the name suggests, it is the low $k$ end
of the integration which is responsible for infrared logarithms. In this
regime $\sin(k \Delta x)/k \Delta x \simeq 1$ and only the first term in the
long wavelength expansion of the mode functions matters,
\begin{equation}
u(t,k) = \frac{H}{\sqrt{2k^3}} \Bigl\{ 1 + \frac12 \Bigl(\frac{k}{H a}
\Bigr)^2 + \frac{i}3 \Bigl(\frac{k}{H a}\Bigr)^3 + \ldots\Bigr\} \; . 
\label{uexp}
\end{equation}

This observation has two important consequences:
\begin{itemize}
\item{The leading logarithm result will not be changed if the
free field mode sum is cut off at $k = H a(t)$; and}
\item{The leading logarithm result will not be changed if the
mode function is replaced by its infrared limit.}
\end{itemize}
Together, these simplifications convert the free quantum field to a 
commuting combination of creation and annihilation operators,
\begin{equation}
\varphi_0(t,\vec{x}) \longrightarrow \Phi_0(t,\vec{x}) \equiv 
\int \frac{d^3k}{(2\pi)^3} \theta(Ha - k) \frac{H}{\sqrt{2 k^3}} 
\Bigl\{e^{i \vec{k} \cdot \vec{x}} \alpha(\vec{k}) + e^{-i \vec{k} \cdot 
\vec{x}} \alpha^{\dagger}(\vec{k}) \Bigr\} \; ,
\end{equation}
Because the Green's function involves a commutator of mode functions, its
infrared truncation requires third order terms from (\ref{uexp}),
\begin{eqnarray}
G(x;x') & \longrightarrow & i\theta(t-t') \int \frac{d^3k}{(2\pi)^3} 
e^{i \vec{k} \cdot \Delta \vec{x}} \frac{H^2}{k^3} \Bigl\{ 
\frac{i}3 \Bigl(\frac{k}{H a}\Bigr)^3 - \frac{i}3 \Bigl(\frac{k}{H a'}\Bigr)^3
\Bigr\} \; , \\
& = &\frac{\theta(t-t')}{3 H} \Bigl[\frac1{a^{\prime 3}} -
\frac1{a^3}\Bigr] \delta^3(\vec{x} - \vec{x}') \; .
\end{eqnarray}
The infrared truncated Yang-Feldman equation is accordingly,
\begin{equation}
\Phi(t,\vec{x}) = \Phi_0(t,\vec{x}) - \frac1{3 H} \int_0^t dt' 
\Bigl[1 - \Bigl(\frac{a'}{a}\Bigr)^3\Bigr] V'\Bigl(\Phi(t',\vec{x})\Bigr) 
\; . \label{almost}
\end{equation}
Although $\varphi(t,\vec{x})$ and $\Phi(t,\vec{x})$ are vastly different
operators, expectation values involving them agree at leading logarithm order.

If we assume the scalars in $V'\Bigl(\varphi(t',\vec{x})\Bigr)$ grow like
$\ln(a')$ --- which is certainly true whenever VEV's are taken --- then
one sees that the $t'$ integration in (\ref{almost}) can produce an additional 
infrared logarithm for the first of the square-bracketed terms. However, the 
rapid growth of the term $(a'/a)^3$ weights the integral overwhelmingly at 
its upper limit and precludes the development of an additional infrared 
logarithm. We can therefore ignore this term and simplify to the equation,
\begin{equation}
\Phi(t,\vec{x}) = \Phi_0(t,\vec{x}) - \frac1{3 H} \int_0^t dt' 
V'\Bigl(\Phi(t',\vec{x})\Bigr) \; .
\end{equation}
Taking the time derivative gives Starobinski\u{\i}'s Langevin equation
\cite{SY},
\begin{equation}
\dot{\Phi}(t,\vec{x}) = \dot{\Phi}_0(t,\vec{x}) - \frac1{3 H} V'\Bigl(
\Phi(t,\vec{x})\Bigr) \; . \label{Langevin}
\end{equation}
Starobinski\u{\i}'s stochastic noise term is the time derivative of the
infrared truncated free field,
\begin{equation}
\dot{\Phi}_0(t,\vec{x}) = \int \frac{d^3k}{(2\pi)^3} \delta(Ha - k) 
\frac{H^2}{\sqrt{2 k}} \Bigl\{e^{i\vec{k} \cdot \vec{k}} \alpha(\vec{k}) +
e^{-i \vec{k} \cdot \vec{x}} \alpha^{\dagger}(\vec{k})\Bigr\} \; .
\end{equation}
A simple calculation reveals that it behaves like white noise,
\begin{equation}
\Bigl\langle \Omega \Bigl\vert \Phi_0(t,\vec{x}) \Phi_0(t',\vec{x}) 
\Bigr\vert \Omega \Bigr\rangle = \frac{H^3}{4 \pi^2} \delta(t - t') \; . 
\label{noise}
\end{equation}

Langevin equations of the form (\ref{Langevin}) have been much studied 
\cite{ALV}. Expectation values of functionals of the stochastic field can 
be computed in terms of a probability density $\varrho(t,\phi)$ as follows,
\begin{equation}
\Big\langle \Omega \Bigl\vert F\left[ \phi(t, {\vec x})\right]
\Bigr\vert \Omega \Big\rangle = \int_{-\infty}^{+\infty} d\phi
\varrho(t,\phi) F(\phi) \; .
\end{equation}
The probability density satisfies a Fokker-Planck equation whose first 
term is given by the interaction in (\ref{Langevin}) and whose second term 
is fixed by the normalization of the white noise (\ref{noise}):
\begin{equation}
\dot{\varrho}(t,\phi) = \frac{1}{3H} \frac{\partial}{\partial \phi}
\Bigl[ V'(\phi) \varrho(t,\phi) \Bigr] + \frac12 \frac{\partial^2}{\partial 
\phi^2} \Bigl[\frac{H^3}{4 \pi^2} \varrho(t,\phi) \Bigr] \; .
\end{equation}
To recover the nonperturbative late time solution of Starobinski\u{\i} 
and Yokoyama \cite{SY} one makes the ansatz,
\begin{equation}
\lim_{t \rightarrow \infty} \varrho(t,\phi) = \varrho_{\infty}(\phi) \; , 
\end{equation}
because the scalar force should eventually balance the tendency of
inflationary particle production to force the scalar up its potential.
This ansatz results in a first order equation,
\begin{equation}
\frac{d \varrho_{\infty}(\phi)} {\varrho_{\infty}(\phi)} =
- \frac{8 \pi^2}{3 H^4} V'(\phi) d\phi \; .
\end{equation}
The solution is straightforward,
\begin{equation}
\varrho_{\infty}(\phi) = N \exp\left[-\frac{8 \pi^2}{3 H^4} V(\phi)\right] \; .
\end{equation}

\section{More General Scalar Models on de Sitter}

A field which can generate infrared logarithms is called {\it active}. Scalar
potential models of the form (\ref{starform}) possess only active fields. 
However, more general theories can possess {\it passive} fields which are not 
themselves capable of engendering an infrared logarithm. A example of such a 
model is scalar quantum electrodynamics (SQED),
\begin{equation}
\mathcal{L} = -\frac14 F_{\mu\nu} F_{\rho\sigma} g^{\mu\rho} g^{\nu\sigma} \!
\sqrt{-g} - \Bigl(\partial_{\mu} \!-\! i e A_{\mu}\Bigr) \varphi^*
\Bigl(\partial_{\nu} \!+\! i e A_{\nu}\Bigr) \varphi g^{\mu\nu} \! \sqrt{-g} 
- \delta \xi \varphi^* \varphi R \sqrt{-g} -\! \frac{\delta \lambda}4
(\varphi^* \varphi)^2 \!\sqrt{-g} \; .
\end{equation}
In this model the charged scalar is active whereas the photon is passive.

Although passive fields cannot cause infrared logarithms, they can propagate
their effects. That is, an expectation value of passive fields can acquire 
an infrared logarithm from a loop correction involving an active field.
For example, the diagram in Fig.~1 gives a contribution to $\langle \Omega 
\vert F_{\mu\nu}(x) F_{\rho\sigma}(x)\vert \Omega \rangle$ which acquires 
an infrared logarithm through the scalar loop at the bottom.
\begin{figure}
\centerline{\epsfig{figure=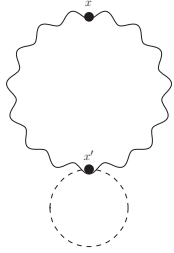,height=2.0in}}
\vskip -2cm
\caption{Two loop contribution to $\langle \Omega \vert F_{\mu\nu}(x) 
F_{\rho\sigma}(x) \vert \Omega \rangle$.}
\end{figure}
Passive fields can also induce interactions between active fields. 
For example, the photon loop in Fig.~2 induces an effective 
$(\varphi^* \varphi)^2$ interaction in SQED.
\begin{figure}
\vskip -2cm
\centerline{\epsfig{figure=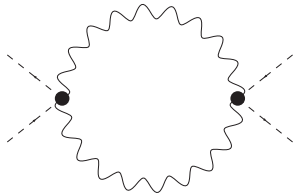,height=3.0in}}
\vskip -3cm
\caption{Effective $(\varphi^* \varphi)^2$ coupling in SQED.}
\end{figure}

In generalizing Starobinski\u{\i}'s technique to theories which include
passive fields, it is crucial to realize that {\it the ultraviolet parts 
of passive fields fields contribute on an equal footing with the infrared 
parts} to the processes of propagating infrared logarithms and mediating 
interactions between active fields. So one cannot infrared truncate the 
passive fields. Instead the correct procedure is:
\begin{itemize}
\item{Integrate out the passive fields and renormalize the resulting
effective action; then}
\item{Infrared truncate and stochastically simplify the purely 
active field effective action.}
\end{itemize}
One might suspect that the second step is not possible owing to the nonlocality
of the effective action. However, first note that VEV's have no spatial 
dependence at leading logarithm order, so one can evaluate all fields at the
same space point. It is still necessary to confront the prospect of fields 
buried inside different temporal integrals. However, {\it precisely because 
they derive from passive fields}, these temporal integrations always contain 
positive powers of the scale factor whose rapid time dependence weights the 
integral overwhelmingly at its upper limit and totally dominates the 
logarithms which might derive from the active fields.  

As an example, consider the retarded Green's function of a conformally
coupled scalar, 
\begin{equation}
G_{\rm cf}(x;x') = \frac{H^2}{4\pi^2} \theta(\Delta t) \frac{\delta(H 
\Delta x + \frac1{a} - \frac1{a'})}{a a' H \Delta x} \; .
\end{equation}
Acting this Green's function on $\ln(a')$ gives,
\begin{eqnarray}
\int_0^t dt' a^{\prime 3} \int d^3x' G_{\rm cf}(x;x') 
\times \ln(a') & = & \frac1{H} \int_0^t dt' \Bigl[\frac{a'}{a} - 
\Bigl(\frac{a'}{a}\Bigr)^2\Bigr] \ln(a') \; , \label{Gcf} \\
& = & \frac1{2 H^2} \Bigl\{\ln(a) - \frac32 + \frac2{a} - \frac1{2a^2} \Bigr\} 
\; .
\end{eqnarray}
(To appreciate the distinction between the Green's functions of passive and
active fields, contrast (\ref{Gcf}) with (\ref{almost}).) On the other hand, 
the result of simply {\it multiplying} $\ln(a)$ by the Green's function acted 
upon unity is,
\begin{equation}
\ln(a) \times \int_0^t dt' a^{\prime 3} \int d^3x' G_{\rm cf}(x;x') 
= \frac{\ln(a)}{2 H^2} \Bigl\{1 - \frac2{a} + \frac1{a^2}\Bigr\} \; . 
\end{equation}
These expressions agree at leading logarithm order, so one may as well locate
all the scalar fields at the same point and evaluate the inverse differential
operators on unity. But that is just the same thing as computing the effective
potential! Hence the hopelessly complicated ``effective action'' degenerates, 
in the leading log approximation, to a very tractable ``effective potential,'' 
and the resulting local theory assumes the form (\ref{starform}) already 
solved by Starobinski\u{\i} \cite{AAS1,SY}.

This program has been carried out for SQED in collaboration with Nikolaos
Tsamis and Tomislav Prokopec \cite{PTsW1}. The various results are best 
expressed in terms of the following quantity,
\begin{equation}
z \equiv \frac{e^2 \varphi^* \varphi}{H^2} \; .
\end{equation}
If one renormalizes to make the quadratic and quartic terms of the 
effective potential vanish, the result is,
\begin{equation}
V_{\rm eff} = \frac{3 H^4}{8 \pi^2} \Biggl\{ \Bigl(-1 \!+\! 2 \gamma\Bigr) z
+ \Bigl(-\frac32 \!+\! \gamma\Bigr) z^2 + \int_0^z dx (1 \!+\! x) \Bigl[
\psi(2  \!-\! \Delta \nu_0) + \psi(1 \!+\! \Delta \nu_0)\Bigr] \Biggr\} \; . 
\label{VSQED}
\end{equation}
Here the function $\Delta \nu_0(x)$ is,
\begin{equation}
\Delta \nu_0(x) \equiv \frac12 - \frac12 \sqrt{1 \!-\! 8 x} = 2x + 4 x^2 +
8 x^3 + O(x^4) \; ,
\end{equation}
and the PolyGamma function is,
\begin{equation}
\psi(1 \!+\! z) \equiv \frac{d}{dz} \ln\Bigl(\Gamma(1 \!+\!z)\Bigr) =
-\gamma + \sum_{n=2}^{\infty} (-1)^n \zeta(n) z^{n-1} \; .
\end{equation}
The effective potential of SQED in de Sitter background does not seem to have 
been computed previously. However, the $H=0$ limit of expression (\ref{VSQED}) 
agrees with equation (4.5) of Coleman and Weinberg \cite{CW}.

To evaluate the VEV of any operator, one first integrates out the passive 
fields and stochastically simplifies the resulting, purely active field
functional. One then computes the VEV using Starobinski\u{\i}'s formalism.
Because the ultraviolet contributes for passive fields, the VEV's of some
operators are ultraviolet divergent even at leading logarithm order. The
scalar functional resulting from the $D$-dimensionally regulated field 
strength bilinear is,
\begin{equation}
F_{\mu\nu}(x) F_{\rho\sigma}(x) \longrightarrow \Bigl( g_{\mu\rho} g_{\nu 
\sigma} - g_{\mu\sigma} g_{\nu\rho}\Bigr) 
\times \frac{H^D}{(4\pi)^{\frac{D}2}} \frac{\Gamma(\frac{D}2 \!-\!
1) \Gamma(2 \!-\! \frac{D}2)}{\Gamma(\frac{D}2 \!+\! 1)} \frac{\Gamma(
\frac{D+1}2 \!+\! \nu) \Gamma(\frac{D+1}2 \!-\! \nu)}{\Gamma(\frac12 \!+\! \nu)
\Gamma(\frac12 \!-\! \nu)} \; . \label{Fexp}
\end{equation}
Here the scalar-dependent parameter $\nu(z)$ is,
\begin{equation}
\nu(z) \equiv \sqrt{\Bigl(\frac{D-3}2\Bigr)^2 - 2 z} \equiv
\Bigl(\frac{D-3}2\Bigr) - \Delta \nu(z) \; .
\end{equation}
The analogous result for the scalar kinetic term is,
\begin{eqnarray}
\lefteqn{\Bigl(\partial_{\mu} \varphi^*(x) - i e A_{\mu}(x) \varphi^*(x) 
\Bigr) \Bigl(\partial_{\nu} \varphi(x) + i e A_{\nu}(x) \varphi(x)\Bigr) }
\nonumber \\
& & \longrightarrow -\frac14 (D\!-\!1) g_{\mu\nu} \times \frac{H^D}{(4\pi)^{
\frac{D}2}} \Biggl\{\frac{\Gamma(\frac{D}2 \!-\! 1) \Gamma(2 \!-\! 
\frac{D}2)}{\Gamma(\frac{D}2 \!+\! 1)} \frac{\Gamma(\frac{D+1}2 \!+\! \nu) 
\Gamma(\frac{D+1}2 \!-\! \nu)}{\Gamma(\frac12 \!+\! \nu) \Gamma(\frac12 \!-\! 
\nu)} - \frac{\Gamma(D\!-\!1)}{\Gamma(\frac{D}2 \!+\!1)} \Biggr\} . 
\qquad \label{phiexp}
\end{eqnarray}
The leading logarithm result for the stress tensor is finite and takes the 
form $-g_{\mu\nu} V_s(z)$ where,
\begin{equation}
V_{\rm s} = \frac{3 H^4}{16 \pi^2} \Biggl\{ \Bigl(-2 \!+\! 2 \gamma\Bigr) z +
\Bigl(-\frac72 \!+\! 2\gamma\Bigr) z^2 + z (1 \!+\! z) \Bigl[\psi(1 \!-\!
2 \Delta \nu_0) + \psi(1 \!+\! \Delta \nu_0)\Bigr] \Biggr\} . \label{stress}
\end{equation}
The stochastic prediction for (\ref{phiexp}) has been checked by an explicit
two loop computation \cite{PTsW2}. The predictions for (\ref{phiexp}) and 
(\ref{stress}) are still being checked.

In collaboration with Shun-Pei Miao, Starobinski\u{\i}'s formalism has also 
been applied to a massless, minimally coupled scalar which is Yukawa-coupled 
to a massless fermion \cite{MW1}. The Lagrangian of this model is,
\begin{equation}
\mathcal{L} = -\frac12 \partial_{\alpha} \varphi
\partial_{\beta} \varphi g^{\alpha\beta} \sqrt{-g} - \frac12 \delta \xi 
\varphi^2 R \sqrt{-g} - \frac1{4!} \delta \lambda \varphi^4 \sqrt{-g}
+ i \overline{\psi} e^{\beta}_{~b} \gamma^b 
{\cal D}_{\beta} \psi \sqrt{-g} - f \varphi
\overline{\psi} \psi \sqrt{-g} \; ,
\end{equation}
where the spinor covariant derivative is,
\begin{equation}
{\cal D}_{\mu} \equiv \partial_{\mu} + \frac{i}2 A_{\mu cd} J^{cd} \; .
\end{equation}
As with SQED, the conformal and quartic counterterms can be chosen to make
the quadratic and quartic terms in the renormalized effective potential 
vanish. The final result is,
\begin{eqnarray}
V_{\rm eff} & = & -\frac{H^4}{4 \pi^2} \sum_{n=2}^{\infty} 
\frac{(-1)^n}{n \!+\! 1} \Bigl[ \zeta(2n \!-\! 1) - \zeta(2n \!+\! 1)\Bigr] 
\Bigl(\frac{f \varphi}{H} \Bigr)^{2n+2} \; , \\
& = & -\frac{H^4}{8 \pi^2} \Biggl\{2 \gamma \Bigl(\frac{f\varphi}{H}\Bigr)^2
- [\zeta(3) \!-\! \gamma] \Bigl(\frac{f\varphi}{H}\Bigr)^4 
+ 2 \int_0^{\frac{f\varphi}{H}} \!\!\!\!\! dx \, (x \!+\! 
x^3) \Bigl[\psi(1 \!+\! i x) \!+\! \psi(1 \!-\! i x)\Bigr] \Biggr\} . 
\qquad \label{VYuk}
\end{eqnarray}
This agrees with the classic result of Candelas and Raine \cite{CR,IMM,IMO}, 
and of course the $H = 0$ limit agrees with equation (6.10) of Coleman and 
Weinberg \cite{CW,BG}. Note that this potential is unbounded below, which 
establishes that infrared logarithms need not always sum up to approach a 
static limit at late times.

As was the case for SQED, the leading logarithm result for the Yukawa stress 
tensor takes the form, $-g_{\mu\nu} V_{\rm s}(\varphi)$, where
\begin{eqnarray}
V_{\rm s}(\varphi) & = & \frac{H^4}{8 \pi^2} \Biggl\{ \Bigl[\frac12 
\!-\! \gamma\Bigr] \Bigl(\frac{f \varphi}{H}\Bigr)^2 + \Bigl[\frac14 \!-\! 
\gamma \!+\! \zeta(3) \Bigr] \Bigl(\frac{f \varphi}{H}\Bigr)^4 \cr
& & \hskip 3cm - \frac12 \Bigl[\Bigl(\frac{f \varphi}{H}\Bigr)^2 \!+\! 
\Bigl(\frac{f \varphi}{H}\Bigr)^4\Bigr] \Bigl[\psi\Bigl(1 \!+\! i\frac{f 
\varphi}{H}\Bigr) + \psi\Bigl(1 \!-\! i\frac{f \varphi}{H}\Bigr)\Bigr] 
\Biggr\} , \label{VsYuk} \\
& = & \frac{H^4}{8 \pi^2} \Biggl\{\frac12 \Bigl(\frac{f 
\varphi}{H}\Bigr)^2 \!+\! \frac14 \Bigl(\frac{f \varphi}{H}\Bigr)^4 
\!-\! \sum_{n=2}^{\infty} (-1)^n \Bigl[\zeta(2n \!-\! 1) \!-\!  \zeta(2n 
\!+\!  1)\Bigr] \Bigl(\frac{f \varphi}{H} \Bigr)^{2n + 2} \Biggr\} .
\end{eqnarray}
The leading logarithm result for the dimensionally regulated coincident 
vertex function is,
\begin{equation}
\varphi(x) \overline{\psi}(x) \psi(x) \longrightarrow 
-4 f \varphi^2(x) \frac{H^{D-2}}{(4 \pi)^{\frac{D}2}} \Biggl\Vert \frac{\Gamma(
\frac{D}2 + \frac{i f \varphi}{H})}{\Gamma(1 + \frac{i f \varphi}{H})} 
\Biggr\Vert^2 \Gamma\Bigl(1 \!-\! \frac{D}2\Bigr) \; . \label{yukcheck}
\end{equation}
Expression (\ref{yukcheck}) was checked at order $f$ against an explicit two 
loop computation \cite{MW1}.

\section{Other Geometries}

Applying Starobinski\u{\i}'s formalism to more general theories on de Sitter
background is essential to resolve the issue of what happens when infrared
logarithms in these models become nonperturbatively strong. However, this 
does not suffice for extrapolating the enhanced quantum effects of these 
models to post-inflationary cosmology. For that it is necessary to understand 
how the effects manifest for a general scale factor $a(t)$. Two issues are
of special importance:
\begin{itemize}
\item{What becomes of the dependence upon the inflationary 
Hubble parameter $H$ which arises from integrating out passive fields?}
\item{What becomes of the infrared logarithms generated by active fields?}
\end{itemize}

Because the passive fields give local effects, it seems plausible that factors 
of $H$ generalize to local curvature scalars. The unique curvature scalar 
of dimension two is the Ricci scalar. Evaluating it respectively for a 
general scale factor and for de Sitter gives,
\begin{equation}
R_{\scriptscriptstyle {\rm FRW}} = 12 H^2(t) + 6 \dot{H}(t) \longrightarrow 
12 H^2 \quad ({\rm de\ Sitter}) \; .
\end{equation}
This suggests that factors of $H^2$ in de Sitter generalize to $R/12$. For
example, the generalizations of the effective potentials of SQED (\ref{VSQED}) 
and Yukawa (\ref{VYuk}) would take the form,
\begin{eqnarray}
V_{\scriptscriptstyle {\rm SQED}} & = & H^4 F\Bigl(\frac{e^2 
\varphi^* \varphi}{H^2}\Bigr) \longrightarrow (\frac1{12} R)^2 
F\Bigl(\frac{12 e^2 \varphi^* \varphi}{R} \Bigr) \; , \\
V_{\scriptscriptstyle {\rm Yukawa}} & = & H^4 G\Bigl(\frac{f^2 \varphi^2}{H^2}
\Bigr) \longrightarrow (\frac1{12} R)^2 G\Bigl(\frac{12 f^2 \varphi^2}{R}\Bigr) 
\; .
\end{eqnarray}
It is presumably this additional dependence upon the metric which is 
responsible for the fact that the potentials of the leading logarithm
stress tensors --- expressions (\ref{stress}) and (\ref{VsYuk}) --- do not 
agree with the effective potentials (\ref{VSQED}) and (\ref{VYuk}). Note 
also that the putative generalization produces a curious sort of $R^2 
f(\varphi^2/R)$ model which may itself have cosmological significance.

Even in de Sitter, infrared logarithms measure the time since the onset of
inflation, so they cannot generalize to a local invariant. One way of 
inferring the nonlocal invariant to which they generalize is to consider the 
expectation value of the square of a free, massless, and minimally coupled 
scalar for an arbitrary scale factor. Because the result is ultraviolet 
divergent, it must be regulated, although the generalized infrared logarithm 
should be finite. If the general $a(t)$ mode function is denoted $u[a](t,k)$, 
the result can be written as follows,
\begin{equation}
\Bigl\langle \Omega_0 \Bigl\vert \varphi_0^2(t,\vec{x}) \Bigr\vert \Omega_0
\Bigr\rangle = \int \frac{d^{D-1}k}{(2\pi)^{D-1}} \Bigl\Vert u[a](t,k)
\Bigr\Vert^2 = \frac2{(4 \pi)^{\frac{D-1}2} \Gamma(\frac{D-1}2)}
\int_H^{\infty} dk k^{D-2} \Bigl\Vert u[a](t,k)\Bigr\Vert^2 \; . \label{log}
\end{equation}
Explicit expressions for $u[a](t,k)$ exist \cite{TW1,TW5}. 
Although these expressions are very complicated, the leading time dependence 
of (\ref{log}) can perhaps be extracted.

\section{Discussion}

There is no question that infrared logarithms arise in explicit perturbative
computations in many inflationary quantum field theories which involve MMC
scalars and/or gravitons 
\cite{PTW1,PTW2,PW1,PW2,GP,OW1,OW2,BOW,TW2,TW3,SW1,MW2,SW2,MW3}. These infrared
logarithms counteract the small coupling constants which would otherwise 
suppress quantum loop effects. Over a long period of inflation they become
so large that weak field perturbation theory breaks down. When the series of
leading infrared logarithms is summed using Starobinski\u{\i}'s formalism,
the result is that MMC scalars always reach nonperturbatively large field
strengths. In the case of Yukawa theory the scalar grows without bound and 
comes to dominate late time cosmology \cite{MW1}. It is not yet known what the 
nonperturbative outcome is for quantum gravity.

These enhanced quantum effects are not restricted to the MMC scalars and 
gravitons which cause them. Other fields seem to experience the following 
effects:
\begin{itemize}
\item{Scalars with nonderivative couplings induce growing masses 
\cite{PTW1,PTW2,PW1,PW2,GP,BOW}; and}
\item{Gravitons and scalars with derivative couplings seem to induce changing 
field strengths \cite{SW1,MW3}.}
\end{itemize}

The now-established fact of significant, nonperturbative quantum 
effects from MMC scalars, and the prospect for them from gravitons, change
the inflationary paradigm. Four processes occur which are absent in classical 
inflation:
\begin{itemize}
\item{Certain field strengths reach nonperturbatively large values 
($\sim H/{\rm coupling\ const}$);}
\item{Inflation-induced masses engender a significant amount of vacuum 
energy which is positive for vector bosons and negative for fermions;}
\item{The effective action of gravity suffers modifications of the form 
$\Delta {\cal L} = R^2 f(\varphi^2/R)$; and}
\item{A potentially significant amount of negative vacuum energy derives 
from the self-gravitation of inflationary gravitons.}
\end{itemize}
Each of these processes can produce observable signals:
\begin{itemize}
\item{Evolving scalars can induce resonant particle production \cite{CKRT};}
\item{Vaccum energy from (Hubble scale) particle masses can serve as a 
reservoir for the creation of a vast population of cosmological wavelength 
vectors \cite{DDPT,DPTD,PW3} and fermions when the mass dissipates after 
second horizon crossing;}
\item{Modifications of gravity can change cosmology \cite{RPW3,NO}; and}
\item{It is conceivable that inflation was caused by a bare, GUT-scale 
cosmological constant which is being screened by infrared quantum
gravitational effects that became strong during primordial inflation 
\cite{TW6}.}
\end{itemize}

\ack The report covers work done jointly with Shun-Pei Miao, Tomislav Prokopec
and Nikolaos Tsamis. I am also grateful for discussions and correspondence 
with Bjoern Garbrecht and Pei-Ming Ho. This work was partially supported by 
NSF grant PHY-0244714 and by the Institute for Fundamental Theory at the 
University of Florida.


\section*{References}
\begin{thebibliography}{99}

\bibitem{RPW1} R. P. Woodard, ``Fermion Self-Energy during Inflation,'' in
{\it XIIth International Conference on Selected Problems of Modern Physics},
edited by B. M. Barbashov, G. V. Efimov, A. V. Efremov, S. M. Eliseev, V. V.
Nesterenko and M. K. Volkov (JINR, Dubna, 2003), pp. 355-366, astro-ph/0307269.

\bibitem{TW1} N. C. Tsamis and R. P. Woodard, Class. Quant. Grav. \textbf{20}
(2003) 5205, astro-ph/\-0206010.

\bibitem{MC} V. F. Mukhanov and G. V. Chibisov, JETP Letters \textbf{33} 
(1981) 532.

\bibitem{AAS2} A. A. Starobinski\u{\i}, JETP Letters \textbf{30} (1979) 682.

\bibitem{PTW1} T. Prokopec, O. Tornkvist and R. P. Woodard, Phys. Rev. Lett.
\textbf{89} (2002) 101301, astro-ph/0205331.

\bibitem{PTW2} T. Prokopec, O. Tornkvist and R. P. Woodard, Ann. Phys. 
\textbf{303} (2003) 251, gr-qc/0205130.

\bibitem{PW1} T. Prokopec and R. P. Woodard, Ann. Phys. \textbf{312} (2204)
1, gr-qc/0310056.

\bibitem{PW2} T. Prokopec and R. P. Woodard, JHEP \textbf{0310} (2003) 059,
astro-ph/0309593.

\bibitem{GP} B. Garbrecht and T. Prokopec, Phys. Rev. D \textbf{73} (2006)
064036, gr-qc/0602011.

\bibitem{OW1} V. K. Onemli and R. P. Woodard, Class. Quant. Grav. \textbf{19}
(2002) 4607, gr-qc/0204065.

\bibitem{OW2} V. K. Onemli and R. P. Woodard, Phys. Rev. D \textbf{70} (2004)
107301, gr-qc/0406098.

\bibitem{VF} A. Vilenkin and L. H. Ford, Phys. Rev. D \textbf{26} (1982) 1231.

\bibitem{ADL} A. D. Linde, Phys. Lett. B \textbf{116} (1982) 335.

\bibitem{AAS3} A. A. Starobinski\u{\i}, Phys. Lett. B \textbf{117} (1982) 175.

\bibitem{BOW} T. Brunier, V. K. Onemli and R. P. Woodard, 
Class. Quant. Grav. \textbf{22} (2005) 59, gr-qc/0408080.

\bibitem{TW2} N. C. Tsamis and R. P. Woodard, Phys. Rev. D \textbf{54} (1996)
2621, hep-ph/9602317.

\bibitem{TW3} N. C. Tsamis and R. P. Woodard, Ann. Phys.
\textbf{253} (1997) 1, hep-ph/9602316.

\bibitem{SW1} S. Weinberg, Phys. Rev. D \textbf{72} (2005) 043514, 2005, 
hep-th/0506236.

\bibitem{Sloth} M. Sloth, Nucl. Phys. B \textbf{748} (2006) 149, 
astro-ph/0604488.

\bibitem{SW2} S. Weinberg, Phys. Rev. {\bf D74} (2006) 023508, hep-th/0605244.

\bibitem{MW2} S. P. Miao and R. P. Woodard, Class. Quant. Grav. \textbf{23}
(2006) 1721, gr-qc/0511140.

\bibitem{MW3} S. P. Miao and R. P. Woodard, Phys. Rev. D \textbf{74} (2006)
024021, gr-qc/0603135.

\bibitem{AAS1} A. A. Starobinski\u{\i}, ``Stochastic de Sitter (inflationary) 
stage in the early universe,'' in {\it Field Theory, Quantum Gravity and
Strings}, edited by H. J. de Vega and N. Sanchez (Springer-Verlag, Berlin, 
1986), pp. 107-126.

\bibitem{SY} A. A. Starobinski\u{\i} and J. Yokoyama, Phys. Rev. D
\textbf{50} (1994) 6357, astro-ph/9407016.

\bibitem{RPW2} R. P. Woodard, Nucl. Phys. Proc. Suppl. \textbf{148} (2005)
108, astro-ph/\-0502556.

\bibitem{TW4} N. C. Tsamis and R. P. Woodard, Nucl. Phys. B \textbf{724}
(2005) 295, gr-qc/0505115.

\bibitem{YF} C. N. Yang and D. Feldman, Phys. Rev. \textbf{79} (1950) 972.

\bibitem{ALV} L. Accardi, Y. G. Lu and I. Volovich, {\it Quantum Theory and 
Its Stochastic Limit} (Springer-Verlag, Berlin, 2002).

\bibitem{PTsW1} T. Prokopec, N. C. Tsamis and R. P. Woodard, Annals Phys.
{\bf 323} (2008) 1324, arXiv:0707.0847.

\bibitem{CW} S. R. Coleman and E. Weinberg, Phys. Rev. D \textbf{7} (1973)
1888.

\bibitem{PTsW2} T. Prokopec, N. C. Tsamis and R. P. Woodard, Class.
Quant. Grav. {\bf 24} (2007) 201, gr-qc/0607094.

\bibitem{MW1} S. P. Miao and R. P. Woodard, Phys. Rev. {\bf D74} (2006)
044019, gr-qc/0602110.

\bibitem{CR} P. Candelas and D. J. Raine, Phys. Rev. D \textbf{12} (1975) 965.

\bibitem{IMM} T. Inagaki, S. Mukaigawa and T. Muta, Phys. Rev. D \textbf{52}
(1995) 4267, hep-th/9505058.

\bibitem{IMO} T. Inagaki, T. Muta and S. D. Odintsov, Prog. Theor. Phys.
Suppl. \textbf{127} (1997) 93, hep-th/9711084.

\bibitem{BG} B. Garbrecht, Phys. Rev. {\bf D74} (2006) 043507,
hep-th/0604166.

\bibitem{TW5} N. C. Tsamis and R. P. Woodard, Class. Quant. Grav. \textbf{21}
(2003) 93, astro-ph/\-0306602.

\bibitem{CKRT} D. J. H. Chung, E. W. Kolb, A. Riotto and I. I. Tkachev, 
Phys. Rev. D \textbf{62} (2000) 043508, hep-ph/9910437.

\bibitem{DDPT} A. C. Davis, K. Dimopoulos, T. Prokopec and O. T\"ornkvist,
Phys. Lett. B \textbf{501} (2001) 165, astro-ph/0007214.

\bibitem{DPTD} K. Dimopoulos, T. Prokopec, O. T\"ornkvist and A. C. Davis,
Phys. Rev. D \textbf{65} (2002) 165, astro-ph/0108093.

\bibitem{PW3} T. Prokopec and R. P. Woodard, Am. J. Phys. \textbf{72} (2004)
60, astro-ph/0303358.

\bibitem{RPW3} R. P. Woodard, Lect. Notes Phys. {\bf 720} (2007) 403,
astro-ph/060167.

\bibitem{NO} S. Nojiri and S. D. Odintsov, Int. J. Geom. Meth. Mod. Phys.
{\bf 4} (2007) 115,
hep-th/0601213.

\bibitem{TW6} N. C. Tsamis and R. P. Woodard, Nucl.Phys. B \textbf{474} (1996)
235, hep-ph/9602315.

\end{thebibliography}
\end{document}